\documentclass[11pt,a4paper]{article}
\usepackage{jheppub_kim}

\usepackage{pdflscape}
\usepackage{amsmath}
\usepackage{amssymb}
\usepackage{dcolumn}
\usepackage{bm}
\usepackage{color}
\usepackage{epsfig}
\usepackage{amsfonts}
\usepackage{graphicx}
\usepackage{epstopdf}
\usepackage{subfigure}
\usepackage{dcolumn}

\newcommand{\be}{\begin{equation}}
\newcommand{\ee}{\end{equation}}
\newcommand{\bea}{\begin{eqnarray}}
\newcommand{\eea}{\end{eqnarray}}

\def\be{\begin{equation}}
\def\ee{\end{equation}}
\def\bea{\begin{eqnarray}}
\def\eea{\end{eqnarray}}

\begin{document}

\title{Constructing an Inflaton Potential by Mimicking Modified Chaplygin Gas}

\author[a]{E.O. Kahya,}

\author[b]{B. Pourhassan,}

\author[a]{S. Uraz}

\affiliation[a]{Physics Department, Istanbul Technical University, Istanbul,
Turkey}

\affiliation[b]{School of Physics, Damghan University, Damghan,
Iran}

\emailAdd{eokahya@itu.edu.tr}
\emailAdd{b.pourhassan@du.ac.ir}
\emailAdd{sedauraz@itu.edu.tr}

\abstract{In this paper, we considered an inflationary model that effectively behaves as a modified Chaplygin gas in the context of quintessence cosmology. We reconstructed the inflaton potential bottom-up and using the recent observational data we fixed the free parameters of the model. We showed that the modified Chaplygin gas inspired model is suitable for both the early and the late time acceleration but has shortcomings between the two periods.}


\emph{}
\maketitle

\section{Introduction}
Chaplygin gas can be thought of as a perfect fluid with a bulk viscosity where the bulk viscosity is the sum of powers of density. Early solutions of this type of inflationary and deflationary models were studied in the context string theory\cite{john1,john2}. When the solution of bosonic d-brane in ($d+1$,1) space-time has been studied classically \cite{P1 0003288} the name Chaplygin gas resurfaced in the context of cosmology. One can begin from Nambu-Goto action in the light-cone gauge and solve the momentum constraint to obtain the Chaplygin gas equation. Later, it was found that other forms of matter on the brane, such as Chaplygin gas, can be introduced \cite{P2 0005011}, where a supersymmetric generalization is only possible for a fluid of the Chaplygin gas type  \cite{P3 0004083}. Friedmann-Robertson-Walker (FRW) cosmological model with Chaplygin gas was proposed as an alternative to quintessence \cite{P4 0103004}.
The Chaplygin gas equation of state,
\begin{equation}\label{I1}
p=-\frac{B}{\rho},
\end{equation}
leads to a homogenous cosmology with,
\begin{equation}\label{I2}
\rho(a)=\sqrt{B+\frac{C}{a^{6}}},
\end{equation}
where $C$ is an integration constant. Therefore, it behaves like dark matter in the early universe $(a\rightarrow0)$ and like dark energy in the late universe $(a\rightarrow\infty)$. This realization led to a proposal \cite{P4 0103004} of inhomogeneous Chaplygin gas which can unify dark matter and dark energy \cite{P5 0111325}. The density perturbations and the mass power spectrum for a universe dominated by the Chaplygin gas have also been studied \cite{P6 0103083,P7 0203441}. The supernova type Ia observational data have been fitted using a model with the Chaplygin gas together CDM \cite{P9 0207430}. It has been shown that the predicted age of the universe in the context of Chaplygin gas CDM
models is in agreement with the most recent age estimates of globular clusters \cite{P10 0209379}.\\

In order for the universe to shift from a dust dominated era to a de Sitter phase, the Chaplygin gas equation of state was generalized to \cite{P4 0103004},
\begin{equation}\label{I3}
p=-\frac{B}{\rho^{\alpha}},
\end{equation}
with the following scale factor-dependent energy density \cite{P11 0202064},
\begin{equation}\label{I4}
\rho(a)=\left[B+\frac{C}{a^{3(1+\alpha)}}\right]^{\frac{1}{1+\alpha}},
\end{equation}
where $C$ is an integration constant and $\alpha>0$ is a free parameter. This model is called the generalized Chaplygin gas model \cite{P17 0209395}. It is clear that $\alpha=1$ reproduces the pure Chaplygin gas model. Constraints on the generalized Chaplygin gas from supernovae observations obtained for the first time by the Ref. \cite{P18 0209486}. Also, statistical analysis of the 2dFGRS data yields to finding that very small (close to zero) and very large values ($\alpha\gg1$) of the equation-of-state parameter $\alpha$ are preferred \cite{P19 1001.4101}. In the recent work \cite{P27 1403.1718} the non-linear regime of unified dark energy models, using the generalized Chaplygin gas cosmology has been studied. Recent observational constraint on generalized Chaplygin gas model indicated that $B = 0.73^{+0.06}_{-0.06}$, $\alpha=-0.09^{+0.15}_{-0.12}$ at $1\sigma$ level and $B = 0.73^{+0.09}_{-0.09}$, $\alpha=-0.09^{+0.26}_{-0.19}$ at $2\sigma$ level \cite{P28 1004.3365} which agree with the claim of the Ref. \cite{P19 1001.4101} for the preferred small value of $\alpha$ while, interestingly, negative value of $\alpha$ was seen to be possible. Choosing negative $\alpha$ changes nature of the generalized Chaplygin gas to some barotropic matter with $p=\omega\rho^{\beta}$, where $\beta(=-\alpha)>0$.
In the other effort \cite{P29 1204.4798} using Markov Chain Monte Carlo method it is found that
$\alpha = 0.00126^{+0.000970+0.00268}_{-0.00126-0.00126}$ and $B = 0.775^{+0.0161+0.0307}_{-0.0161-0.0338}$. While there are several attempt to fix $B$ using observational data, some authors suggested $B$ as variable parameter \cite{P30 1203.4637,P31 1210.5021} and the resulting model called varying generalized Chaplygin gas or new generalized Chaplygin gas model. There is also the possibility to consider viscosity in the generalized Chaplygin gas model \cite{P36 0511814}.\\
The generalized Chaplygin gas equation of state can be extended to the following form \cite{P42 0305559},
\begin{equation}\label{I5}
p=-\frac{\frac{B}{1+\omega}-1+(\rho^{1+\alpha}+1-\frac{B}{1+\omega})^{-\omega}}{\rho^{\alpha}}
\end{equation}
which is called the generalized cosmic Chaplygin gas with $0 > \omega > -l$, and $l$ being a positive
definite constant which can take on values larger than unity \cite{P43 1112.6154}.
There is also a class of equations of state interpolating between ordinary fluids with high energy densities and generalized Chaplygin gas fluids with low energy densities which is called the modified Chaplygin gas \cite{P47 0205140}. Soon after that viscosity was introduced to the modified Chaplygin gas \cite{P54 0801.2008}. Finally further extensions were introduced as modified cosmic Chaplygin gas \cite{P72,P73 1304.6987,P74}, and extended Chaplygin gas \cite{P75 1402.2592,P76 1405.0667,a,b,c}. The modified Chaplygin gas is subjects of this paper to study inflation.\\
Inflationary scenarios including generalized Chaplygin gas have been considered already \cite{P77 1212.2641,P78 1303.5658,P79 1310.4988}. Recently, the generalized Chaplygin gas inflation in the light of Planck \cite{P80 1303.5075} and BICEP2 \cite{P81 1403.3985} has been studied and has been shown that the generalized Chaplygin gas is not a suitable candidate for inflation in the context of Einstein theory of gravity \cite{P82 1404.3683}. Now, we would like to investigate the consequences of Inflation obeying the equation of state of the modified Chaplygin gas so that it can explain both the primordial and late time acceleration. We will use recent observational data to fix parameters of the modified Chaplygin gas model,  in other words we would like to construct the inflaton potential in order to get a viable mechanism during both of the accelerated stages.\\
The paper is organized as follows. In next section we introduce modified Chaplygin gas inflation and obtain the scale factor dependence of the scalar field and construct the inflaton potential by solving Friedmann equations. Then, in section 3, we calculate the tensor to scalar ratio, slow-roll parameters, scalar spectral index by using the inflaton potential and use Planck data to fix the model parameters. Using fixed parameters we investigate some cosmological parameters in section 4. In section 5 we briefly examine the effects of varying bulk viscosity in modified Chaplygin gas. In section 6 we give our conclusions and suggest some future work.

\section{Modified Chaplygin gas inflation}
Although the Chaplygin gas models were introduced to explain late time acceleration without dark energy, the primordial acceleration which is believed to be driven by a scalar field, Inflaton, can also be described by the same modified Chaplygin gas model. Therefore we would like to describe both the primordial and recent inflationary phase with a scalar field, that we can still call inflaton, for which the equation of state is the one that the modified Chaplygin gas obeys.\\

The MCG is defined with the following equation of state \cite{P83 1211.3518},
\begin{equation}\label{s1}
p=A\rho-\frac{B}{\rho^{\alpha}},
\end{equation}
where $A$, $\alpha$, and $B$ are free parameters of the model. The case of $A = 0$ recovers generalized Chaplygin gas EoS, and $A = 0$ together $\alpha = 1$ recovers the original Chaplygin gas EoS. The special case of $\alpha=0.5$ has been studied in the Ref. \cite{P84 1312.0779}. Ref. \cite{P85 1004.3364} where it was concluded that the best fitted parameters are $A = -0.085$ and $\alpha = 1.724$. Other observational constraints on the modified Chaplygin gas model using Markov Chain Monte Carlo approach resulted into $A = 0.00189^{+0.00583}_{-0.00756}$, $\alpha=0.1079^{+0.3397}_{-0.2539}$ at $1\sigma$ level and $A = 0.00189^{+0.00660}_{-0.00915}$, $\alpha=0.1079^{+0.4678}_{-0.2911}$ at $2\sigma$ level \cite{P87 1105.1870}. Other interesting constraints may be also be found based on cosmic growth \cite{P91 1306.4808}.\\

All of the observational constraints given above can be summarized with a single statement; parameter $A$ may be a small positive or small negative number. Many of these analysis were done with a very strong simplification; Chaplygin gas being the only matter in the universe. It is relatively safe to make that simplification when considering a non-zero parameter $B$; since for the final stages of the universe, with decreasing energy density, Chaplygin gas is going to dominate the universe and will determine the Hubble parameter.

We would like to constrain the parameter $A$ by assuming that the same scalar is causing the primordial inflation unlike the studies above. Primordial inflation is described by quasi de Sitter geometry with the equation of state being $w = -1 + \epsilon$ where $\epsilon$ is a positive infinitesimally small number. Therefore, one can trivially infer from equation (\ref{s1}) that parameter $A$ should be very close to -1 since that part of the pressure will dominate during the early universe. And the peculiar form of equation (\ref{s1}) is going to give rise to a specific inflaton potential which we will derive below.\\

Assuming flat FRW universe, and taking $8 \pi G =1 $ yields to the following Friedmann equation,
\begin{equation}\label{s2}
H^{2}=\left(\frac{\dot{a}}{a}\right)^{2}=\frac{\rho}{3},
\end{equation}
where $H$ is the Hubble parameter and $a$ is the scale factor. Also, one can obtain conservation equation as
\begin{equation}\label{s3}
\dot{\rho}+3H(p+\rho)=0.
\end{equation}

Using the relations (\ref{s1}) and (\ref{s2}) in the equation (\ref{s3}) gives the following scale factor-dependence for energy density \cite{P93 1205.3768},
\begin{equation}\label{s4}
\rho=\left(\frac{B}{1+A}+\frac{C}{a^{3(1+A)(1+\alpha)}}\right)^{\frac{1}{1+\alpha}},
\end{equation}
where $C$ is an integration constant.\\
It is clear from the equation (\ref{s1}) that during the early universe $p\simeq A\rho$. And for the late time, one can see from the equation (\ref{s4}) that energy density approaches to a constant $\rho\rightarrow(\frac{B}{1+A})^{\frac{1}{1+\alpha}}$ since $a\gg1$. If we put this value for $\rho$ in the equation (\ref{s1}) for pressure;
\begin{equation}\label{s5}
p=A(\frac{B}{1+A})^{\frac{1}{1+\alpha}}-B(\frac{B}{1+A})^{-\frac{\alpha}{1+\alpha}}=-(\frac{B}{1+A})^{\frac{1}{1+\alpha}}=-\rho,
\end{equation}
The above equation tells us that the modified Chaplygin gas behaves like cosmological constant during the later stage of expansion independent of parameters $A$, $B$ or $\alpha$.

Let us assume that the quintessence scalar field $\phi$ is minimally coupled to gravity \cite{P94 1310.7167}, and it is the scalar field which derives inflation. We would like this scalar field to obey the equation of state of the modified Chaplygin gas. One can achieve this by assuming that energy density and pressure are given by the following relations respectively,
\begin{eqnarray}\label{s6}
\rho&=&\frac{1}{2}\dot{\phi}^{2}+V(\phi),\nonumber\\
p&=&\frac{1}{2}\dot{\phi}^{2}-V(\phi),
\end{eqnarray}
where $V(\phi)$ is the potential for the scalar field. Using equations (\ref{s6}) and (\ref{s1}) one can get $\dot{\phi}$ to have the following form
\begin{equation}\label{s7}
\dot{\phi}=\sqrt{(1+A)\rho-\frac{B}{\rho^{\alpha}}}.
\end{equation}
Now, using the equations (\ref{s2}), (\ref{s4}) and (\ref{s7}) we can obtain,
\begin{equation}\label{s8}
\frac{d\phi}{da}=\frac{\sqrt{3(1+A)}}{a\sqrt{1+\frac{B}{C(1+A)}a^{3(1+A)(1+\alpha)}}}.
\end{equation}
The equation (\ref{s8}) has the following solution,
\begin{equation}\label{s9}
\phi=-\frac{2\sqrt{3(1+A)}}{3(1+A)(1+\alpha)}\cosh^{-1}{\left(\sqrt{1+\frac{(1+A)C}{B}a^{-3(1+A)(1+\alpha)}}\right)}+\phi_{f},
\end{equation}
where $\phi_{f}$ is final value (when $a\gg1$) of the scalar field.
Conversely, we can express the scale factor in terms of the scalar field,
\begin{equation}\label{s10}
a=\left[\frac{B}{(1+A)C}(-1+\cosh^{2}{\theta})\right]^{-\frac{1}{3(1+A)(1+\alpha)}},
\end{equation}
where we defined,
\begin{equation}\label{s11}
\theta\equiv\frac{3(1+A)(1+\alpha)}{2\sqrt{3(1+A)}}(\phi-\phi_{f}).
\end{equation}
Having known the dependence of the the scale factor on the scalar field we can obtain the expressions for the scalar potential and the Hubble expansion parameter as a function of the scalar field. Using the relations (\ref{s4}), (\ref{s6}) and (\ref{s10}) we can find,
\begin{equation}\label{s12}
V=V_{0}\left[1-\frac{A-1}{A+1}\cosh^{2}{\theta}\right]\left(\cosh{\theta}\right)^{-\frac{2\alpha}{1+\alpha}},
\end{equation}
where $V_{0} = \frac{A+1}{2} (\frac{B}{1+A})^{\frac{1}{1+\alpha}}$ is a constant that depends on $A$, $B$ and $\alpha$.
Then, using the relations (\ref{s2}), (\ref{s4}) and (\ref{s10}) one can obtain,
\begin{equation}\label{s14}
H=H_{0}\left(\cosh{\theta}\right)^{\frac{1}{(1+\alpha)}},
\end{equation}
where the constant $H_0$ has a particular dependence on $A$, $B$ and $\alpha$ of the following form, $H_{0} = \frac{1}{\sqrt3} (\frac{B}{1+A})^{\frac{1}{2+2\alpha}}$, which we define for convenience.
In the next section we will use above result to obtain the tensor to scalar ratio and fix the free parameters using Planck 2015 + BICEP2 data. In section 4, having fixed the free parameters of the model, we will analyze the behavior of cosmological parameters such as Hubble expansion parameter and the deceleration parameters.

\section{The tensor to scalar ratio}
\begin{figure}[h!]
 \begin{center}$
 \begin{array}{cccc}
\includegraphics[width=60 mm]{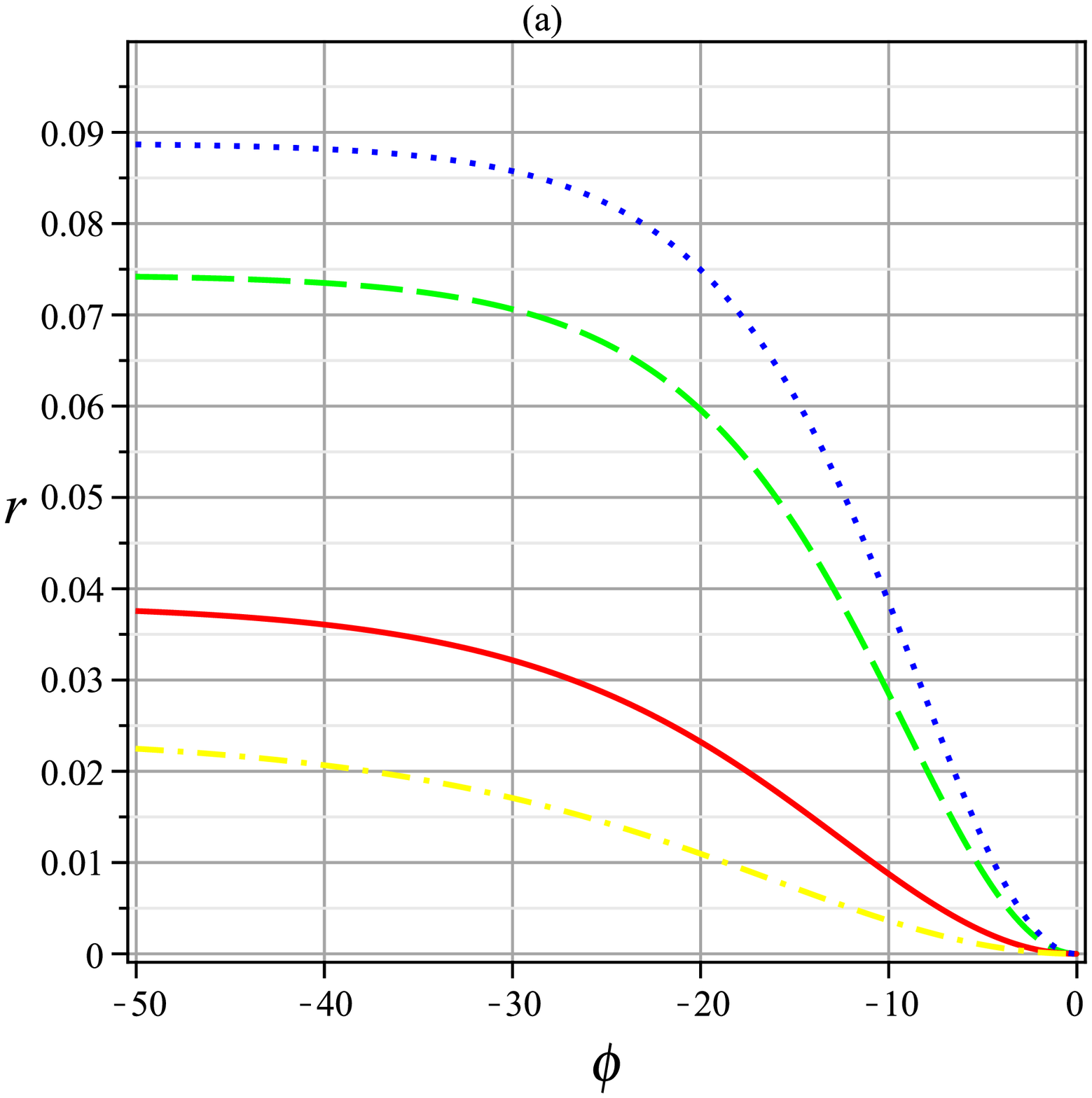}\includegraphics[width=60 mm]{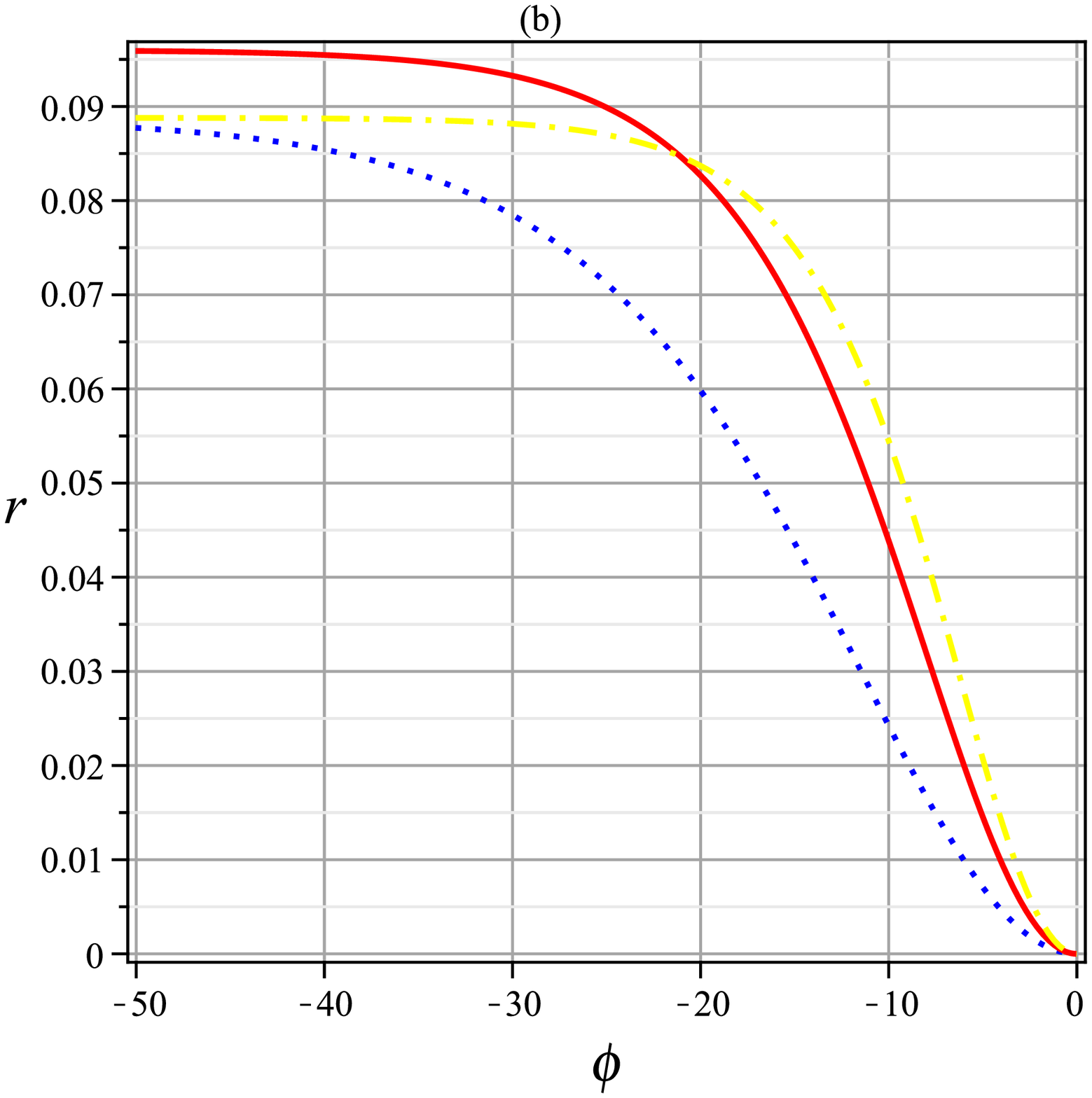}
 \end{array}$
 \end{center}
\caption{The tensor to scalar ratio in terms of the scalar field for $B=3$, $C=1$ and $\phi_{f}=0$. (a) $\alpha=0.5$, $A=-0.996$ (dotted blue), $A=-0.997$ (dashed green), $A=-0.998$ (solid red), $A=-0.999$ (dash dotted yellow). (b) $A=-0.996$, $\alpha=0.1$ (dotted blue), $\alpha=0.5$ (solid red), $\alpha=1$ (dash dotted yellow).}
 \label{fig1}
\end{figure}

The first constraint that we will make use of is the tensor to scalar ratio. Assuming slow-roll inflation $r$, the tensor to scalar ratio which measures the amount of the stochastic gravitational wave production during inflation, is given by,
\begin{equation}\label{s16}
r\simeq16\epsilon_{H}.
\end{equation}
Here $\epsilon_{H}$ is one of the slow-roll parameters which is defined as,
\begin{equation}\label{s17}
\epsilon_{H}=2\left(\frac{\frac{\partial H}{\partial\phi}}{H}\right)^{2}.
\end{equation}
Using the equations (\ref{s14}) and (\ref{s17}) we get the following expression,
\begin{equation}\label{s18}
\epsilon_{H}=\frac{3}{2}(1+A)\tanh^{2}{\theta}.
\end{equation}
Therefore the tensor to scalar ratio can be expressed in terms of the scalar field as
\begin{equation}\label{s19}
r \simeq 24(1+A)\tanh^{2}{\theta}.
\end{equation}
On the other hand, the current upper bound on $r$ by BKP (BICEP Keck Planck \cite{1502.01589}) is $r\leq0.09$. The equation (\ref{s19}) tells us that, in the early universe $\theta\rightarrow-\infty$ then $\tanh^{2}(\theta)\rightarrow1$ and we have $r\rightarrow 24(1+A)$. It means that the asymptotic behavior of the tensor to scalar ratio is $24(1+A)$, hence, the only free parameter that determines the tensor to scalar ratio is $A$; which will be fixed using observational data of BKP as,
\begin{equation}\label{s20}
-1<A<-0.996.
\end{equation}
In the case of upper limit of the BKP joint analysis one gets $A=-0.996$. This is an expected result as we discussed in the introduction section and is different from the previous results. For example, authors of the Ref. \cite{P99 0905.2281} suggested $0\leq A\leq1.35$, also there are several works which used $A\geq0$. Although there are some works which found small negative values for the parameter $A$, at least with our knowledge, the value of $A\approx-0.99$ is a new result. This is due to enforcing the scalar field to be the cause of the early time and late time acceleration. In Fig. \ref{fig1} we can see the behavior of $r$ by various values of $\alpha$. Assuming $\phi$ an increasing function of time, the limit of $r\leq0.09$ is achieved for $A\approx-0.99$. This proves our expectation of the asymptotic behavior of $r$, that it is independent of all parameters other than $A$.\\

\begin{figure}[h!]
 \begin{center}$
 \begin{array}{cccc}
\includegraphics[width=70 mm]{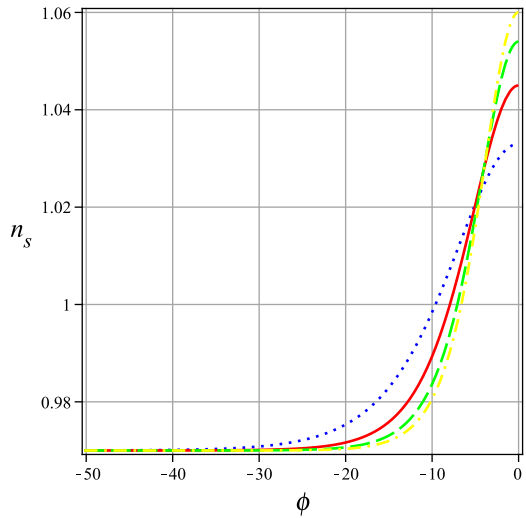}
\end{array}$
 \end{center}
\caption{The scalar spectral index in terms of the scalar field for $\alpha=0.5$, $C=1$ and $\phi_{f}=0$. $A=-0.996$ (dotted blue), $A=-0.997$ (solid red), $A=-0.998$ (dashed green), $A=-0.999$ (dash dotted yellow).}
 \label{fig2}
\end{figure}

Let us look at another important observable quantity, the scalar spectral index $n_{s}$ which is given by,
\begin{equation}\label{s21}
n_{s}\simeq 1-4\epsilon_{H}+2\delta_{H},
\end{equation}
where,
\begin{equation}\label{s22}
\delta_{H}=\epsilon_{H}-\left(\frac{\dot{\epsilon_{H}}}{2H\epsilon_{H}}\right),
\end{equation}
is the other slow-roll parameter. The current bound on $n_{s}$ obtained by Planck 2015 TT + low P analysis \cite{1502.01589} is $n_{s}=0.968\pm0.006$. In Fig. \ref{fig2} we draw $n_{s}$ in terms of the scalar field and see that, at the early universe $\phi\rightarrow-\infty$, the observed upper bound is satisfied.

Having a negative but bigger than -1 value for $A$ would constrain $B$ to be a positive parameter. This can be seen from the following; equation (\ref{s14}) shows us that since $H_{0} = \frac{1}{\sqrt3} (\frac{B}{1+A})^{\frac{1}{2+2\alpha}}$ only a positive value of $B$ gives a positive real Hubble parameter.\\
The usual choice for the value of parameter $\alpha$ is between zero and unity. As an example, which is one among the many, in Ref. \cite{P84 1312.0779} the case of positive $\alpha=1/2$ is considered and the model is called the reduced modified Chaplygin gas model. In late time limit one could get the pressure contribution of Chaplygin gas to dominate the total pressure only if pressure is inversely proportional to density, which is a decreasing function of time. Despite this fact, in Ref. \cite{P82 1404.3683} some cases of negative $\alpha$ were also discussed. The sign and the value of $\alpha$ will be constrained using the current value of the Hubble parameter.
In the next section we study cosmological parameters using obtained parameter values.
\section{Cosmological parameters}
\begin{figure}[h!]
 \begin{center}$
 \begin{array}{cccc}
\includegraphics[width=70 mm]{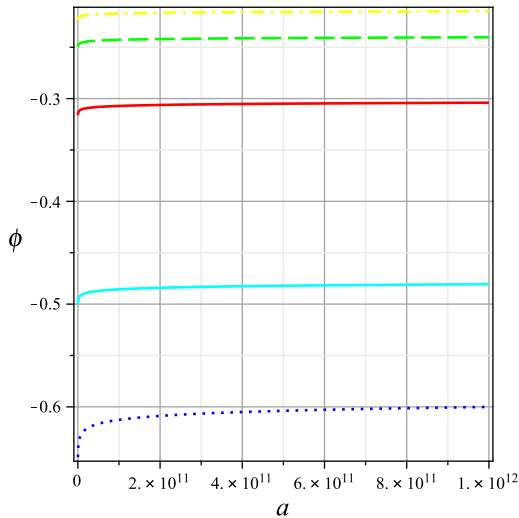}
 \end{array}$
 \end{center}
\caption{The scalar field in terms of the scale factor for $A=-0.996$, $\alpha=0.5$, $C=1$ and $\phi_{f}=0$. $B=1$ (dotted blue), $B=2$ (solid cyan)$B=5$ (solid red), $B=8$ (dashed green), $B=10$ (dash dotted yellow).}
 \label{fig3}
\end{figure}

\begin{figure}[h!]
 \begin{center}$
 \begin{array}{cccc}
\includegraphics[width=70 mm]{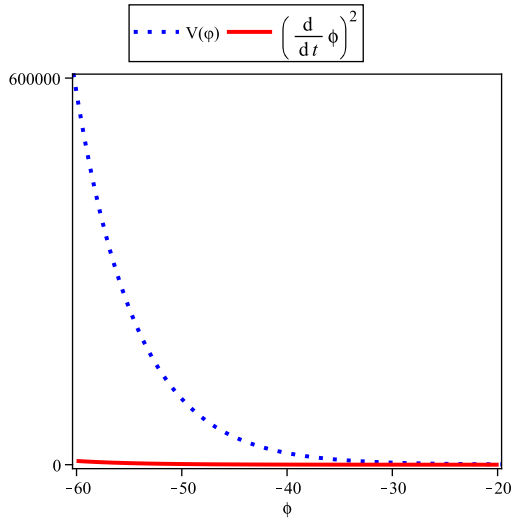}
 \end{array}$
 \end{center}
\caption{$\dot{\phi}^{2}$ (solid red) and $V(\phi)$ (dotted blue) in terms of the scalar field for $A=-0.996$, $\alpha=0.5$, $B=3$, $C=1$ and $\phi_{f}=0$.}
 \label{fig4}
\end{figure}

Using equation (\ref{s9}) we can obtain the behavior of the scalar field with the scale factor as illustrated in Fig. \ref{fig3}. We can see that $\phi$ is an increasing function of the scale factor. With the choice of $\phi_{f}=0$ the value of the scalar filed goes from negative infinity to zero. It is clear from Fig. 3 that the evolution of $\phi$ at the initial states are faster that the later
states. This might be a troublesome situation, since the early universe inflation is achieved by having the scalar field slowly roll from the inflaton potential, in turn $\dot{\phi}^{2}$ to be small
compared to scalar potential. Here we can see from the Fig. \ref{fig4} that the magnitude of $V(\phi)$ is so big compared to $\dot{\phi}^{2}$ that, the $\ddot{\phi}$ doesn't help $\dot{\phi}^{2}$ to catch the value of the inflaton potential during the early universe for a long period of time.

\begin{figure}[h!]
 \begin{center}$
 \begin{array}{cccc}
\includegraphics[width=70 mm]{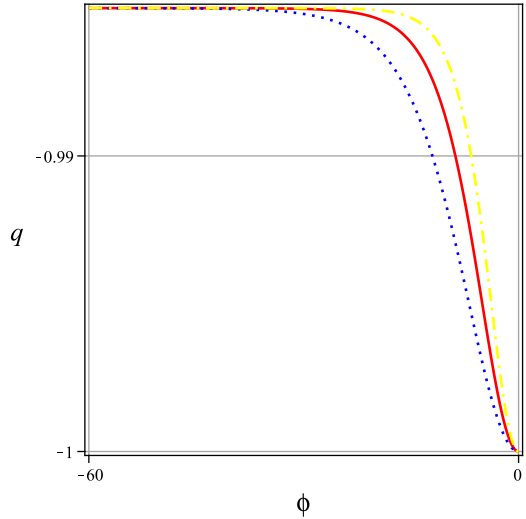}
 \end{array}$
 \end{center}
\caption{The deceleration parameter in terms of the scale factor for $A=-0.996$, $B=3$, $C=1$ and $\phi_{f}=0$.  $\alpha=0.1$ (dotted blue), $\alpha=0.5$ (solid red), $\alpha=1$ (dash dotted yellow).}
 \label{fig5}
\end{figure}

Then, using equation (\ref{s14}) we can investigate the evolution of the Hubble parameter. One can see that the Hubble parameter is a decreasing function of scalar field and yields to a constant at the late times. Using the current value of Hubble parameter ($H\approx67 km s^{-1} Mpc^{-1}\approx10^{-61}$) one can also fix $B$ and $\alpha$. The asymptotic value of Hubble parameter is found in section 2 to be $\frac{1}{\sqrt{3}}(\frac{B}{1+A})^{\frac{1}{2+2\alpha}}$. To make this number close to $10^{-61}$ (Hubble parameter's current value in natural units) one has two choices:
\begin{itemize}
\item $\alpha>0$, $B\sim10^{-100}$ for Chaplygin gas.
\item $\alpha<0$, $B\sim\mathcal{O}(1)$ for barotropic fluid.
\end{itemize}
However they don't have any effect on $r$ and behavior of other cosmological parameters. Having $H$ in terms of $\phi$ one can obtain the deceleration parameter $q$ as,
\begin{equation}\label{sq}
q=-1-\frac{\dot{H}}{H^{2}}=-1-\frac{\sqrt{(1+A)\rho-\frac{B}{\rho^{\alpha}}}}{H^{2}}\frac{dH}{d\phi}.
\end{equation}
So, we represent behavior of $q$ in terms of the scale factor by Fig. \ref{fig5} in exact agreement with observations which tells us that the final value of the deceleration parameter is about -1. We can see that the deceleration parameter initially begins with a negative value and approaches to -1 towards the end (late time). Above results suggest that our model can explain both the early and late time acceleration with a single inflaton potential $V(\phi)$. The problem is ending it during the intermediate stages. One can imagine a scenario where $\phi$ decays into other scalar particles at low energies and give rise to reheating but eventually dominate the universe due to its form (constant energy density).\\
In the next section we include bulk viscosity and discuss about consequences.

\section{Effect of varying bulk viscosity}
In this section we assume that the space-time is filled with modified Chaplygin gas having a bulk viscosity. In that case, the total pressure changes as the following:
\begin{equation}\label{s24}
p_{eff}=p+\Pi.
\end{equation}
Here the bulk viscous pressure $\Pi$ is represented
by the Eckart�s expression which is proportional to the Hubble parameter with proportionality factor identified
as the bulk viscosity coefficient,
\begin{equation}\label{s25}
\Pi=-3\xi H.
\end{equation}
The bulk viscous coefficient is assumed to have a power law dependence on the energy density,
\begin{equation}\label{s26}
\xi=\xi_{0}\rho^{n},
\end{equation}
where $\xi_{0}$ and $n$ are positive constants. Possible solutions were discussed for this proposed model  \cite{P57 1401.8002}. The simplest solution obtained for the special case of $n=1/2$ is considered here to obtain the effect of bulk viscosity on observed quantities such as $n_{s}$ and $r$. In the case of $n=1/2$ one can obtain \cite{P57 1401.8002} the following expression for the energy density:
\begin{equation}\label{s27}
\rho=\left(\frac{B}{1+A-\sqrt{3}\xi_{0}}+\frac{C}{a^{3(1+A-\sqrt{3}\xi_{0})(1+\alpha)}}\right)^{\frac{1}{1+\alpha}}.
\end{equation}
Repeating the similar calculations of the previous section shows us that the presence of viscosity is not suitable for modified Chaplygin gas model and this is in accordance with the suggestion that the modified Chaplygin gas behaving as viscous generalized Chaplygin gas. We find that value of the $n_{s}$ increases due to bulk viscosity, while the tensor to scalar ratio decreases by viscous coefficient. And the problem of getting the initial and final inflation consistently by this modification still remains.

\section{Conclusion and discussion}
In this paper we presented a model in which a single scalar field inflaton is responsible for both the primordial and late time acceleration. We forced this scalar field to obey the modified Chaplygin gas EoS. By doing so, we constructed the inflaton potential bottom-up. In a recent work \cite{1405.7681} a similar approach was used to explain both periods using a k-essence non-minimally coupled scalar field. Unlike giving the form of the potential initially; we wanted the Hubble parameter dictate the form of the inflaton potential by assuming Chaplygin gas EoS via Friedmann equations. Recently \cite{P82 1404.3683}, using Planck+BICEP2 data, it is found that generalized Chaplygin gas is not a suitable model of inflation in Einstein gravity, so one may need generalized Chaplygin gas inflationary model in RS type five dimensional brane world scenario. In this paper, we investigated the same problem for the modified Chaplygin gas model with and without viscosity in the light of more recent data \cite{1502.01589}.
In the modified Chaplygin gas model there are three fundamental parameters: $A$, $B$ and $\alpha$. We obtained that the most relevant free parameter to primordial inflation, $A$ should lie between $-1\leq A\leq-0.996$. The values for $B$ and $\alpha$ are determined by the current value of Hubble parameter. The conclusion is that $\alpha$ is either a positive $\mathcal{O}(1)$ number or a negative number which is very close to -1. The former case gives an unnaturally small value for $B\sim10^{-100}$ or so. The latter case works with $B\sim\mathcal{O}(1)$ which would certainly be preferred. Therefore, one can say that late time acceleration caused by the same inflation field, prefers barotropic fluid EoS over Chaplygin gas EoS. We conclude that the presence of bulk viscosity is not necessary in modified Chaplygin gas and may yield to inconsistencies with observational data.\\
The only shortcoming of our model is a mechanism to end inflation between two accelerated stages. This model which basically behaves quite similar to cosmological constant will obviously not allow production of large scale structure. This feature can be analyzed formally by solving perturbed Einstein's equations but Fig. \ref{fig5} leaves no room for that effort. Ending inflation can be achieved by some coupling to a spectator scalar field and therefore a decay to that particle would soon give rise to reheating. But this ad-hoc procedure should be developed more formally. \\
For future work, it can be considered to use extended models including other free parameters \cite{P42 0305559}. One can actually imagine a density term, with a postive constant in front, in the equation of state. If the power of that density term is negative than it would dominate the pressure between the two accelerated stages. One could solve the Friedmann equations to construct Inflaton potential for this case as well. Cosmological perturbation analysis would help to constrain additional free parameters that will enter with the new term. We can also investigate k-essence generalization and even non-minimal coupling between scalar field and gravity. All of these attempts will complicate the problem but nevertheless will give us more space to explain observable phenomena.

\section{Acknowledgement}
We thank Richard Woodard for helpful discussions. SU also thanks to Ali Kaya and Savas Arapoglu for helpful comments. EOK and BP acknowledge support from Tubitak Grant Number: 112T817.


\begin{thebibliography}{99}
\bibitem{john1}
John D. Barrow, "The deflationary universe: An instability of the de Sitter universe" Phys. Lett. B 180 (1986) 335
\bibitem{john2}
John D. Barrow, "String-driven inflationary and deflationary cosmological models", Nucl. Phys. B 310 (1988) 743
\bibitem{P1 0003288}
N. Ogawa, "A Note on Classical Solution of Chaplygin-gas as D-brane", Phys. Rev. D 62 (2000) 085023
\bibitem{P2 0005011}
A. Yu. Kamenshchik, U. Moschella, V. Pasquier, "Chaplygin-like gas and branes in black hole bulks", Phys. Lett. B487 (2000) 7
\bibitem{P3 0004083}
R. Jackiw and A. P. Polychronakos, "Supersymmetric fluid mechanics", Phys. Rev. D 62 (2000) 085019
\bibitem{P4 0103004}
A. Yu. Kamenshchik, U. Moschella, V. Pasquier, "An alternative to quintessence", Phys. Lett. B 511 (2001) 265
\bibitem{P5 0111325}
N. Bilic, G.B. Tupper, R.D. Viollier, "Unification of Dark Matter and Dark Energy: the Inhomogeneous Chaplygin Gas", Phys. Lett. B 535 (2002) 17
\bibitem{P6 0103083}
J.C. Fabris, S.V.B. Goncalves, P.E. de Souza, "Density perturbations in an Universe dominated by the Chaplygin gas", Gen. Rel. Grav. 34 (2002) 53
\bibitem{P7 0203441}
J. C. Fabris, S.V.B. Gon�alves, P.E. de Souza, "Mass Power Spectrum in a Universe Dominated by the Chaplygin Gas", Gen. Rel. Grav. 34 (2002) 2111
\bibitem{P9 0207430}
J.C. Fabris, S.V.B. Goncalves, P.E. de Souza, "Fitting the Supernova Type Ia Data with the Chaplygin Gas", [arXiv:astro-ph/0207430]
\bibitem{P10 0209379}
A. Dev, J. S. Alcaniz, D. Jain, "Cosmological consequences of a Chaplygin gas dark energy", Phys. Rev. D 67 (2003) 023515
\bibitem{P11 0202064}
M.C. Bento, O. Bertolami, A.A. Sen, "Generalized Chaplygin Gas, Accelerated Expansion and Dark Energy-Matter Unification", Phys. Rev. D 66 (2002) 043507
\bibitem{P17 0209395}
V. Gorini, A. Kamenshchik, U. Moschella, "Can the Chaplygin gas be a plausible model for dark energy?", Phys. Rev. D 67 (2003) 063509
\bibitem{P18 0209486}
M. Makler, S.Q. de Oliveira, I. Waga, "Constraints on the generalized Chaplygin gas from supernovae observations", Phys. Lett. B 555 (2003) 1
\bibitem{P19 1001.4101}
J. C. Fabris, H.E.S. Velten, W. Zimdahl, "Matter power spectrum for the generalized Chaplygin gas model: The relativistic case", Phys. Rev. D 81 (2010) 087303
\bibitem{P27 1403.1718}
P.P. Avelino, K. Bolejko, G.F. Lewis, "Non-linear Chaplygin Gas Cosmologies", [arXiv:1403.1718 [astro-ph.CO]]
\bibitem{P28 1004.3365}
J. Lu, Y. Gui, L. Xu, "Observational constraint on generalized Chaplygin gas model", Eur. Phys. J. C 63 (2009) 349
\bibitem{P29 1204.4798}
L. Xu, J. Lu, Y. Wang, "Revisiting Generalized Chaplygin Gas as a Unified Dark Matter and Dark Energy Model", Eur. Phys. J. C 72 (2012) 1883
\bibitem{P30 1203.4637}
A. Aviles, A. Bastarrachea-Almodovar, L. Campuzano, H. Quevedo, "Extending the generalized Chaplygin gas model by using geometrothermodynamics", Phys. Rev. D 86 (2012) 063508
\bibitem{P31 1210.5021}
K. Liao, Y. Pan, Z-H. Zhu, "Observational constraints on new generalized Chaplygin gas model", Research in Astronomy and Astrophysics 13 (2013) 159
\bibitem{P36 0511814}
X-H. Zhai, Y-D Xu, X-Z Li, "Viscous generalized Chaplygin gas", Int. J. Mod. Phys. D15 (2006) 1151
\bibitem{P42 0305559}
P.F. Gonzalez-Diaz, "You need not be afraid of phantom energy", Phys. Rev. D 68 (2003) 021303
\bibitem{P43 1112.6154}
J. Bhadra, U. Debnath, "Accretion of New Variable Modified Chaplygin Gas and Generalized Cosmic Chaplygin Gas onto Schwarzschild and Kerr-Newman Black holes", Eur. Phys. J. C 72 (2012) 1912
\bibitem{P47 0205140}
H.B. Benaoum, "Accelerated Universe from Modified Chaplygin Gas and Tachyonic Fluid", [arXiv:hep-th/0205140]
\bibitem{P54 0801.2008}
C.S.J. Pun, L.�. Gergely, M.K. Mak, Z.Kov�cs, G.M. Szab�, T. Harko, "Viscous dissipative Chaplygin gas dominated homogenous and isotropic cosmological models",  Phys. Rev. D77 (2008) 063528
\bibitem{P72}
H. Saadat and B. Pourhassan, "FRW bulk viscous cosmology with modified cosmic Chaplygin gas", Astrophys. Space Sci. 344 (2013) 237
\bibitem{P73 1304.6987}
J. Sadeghi, H. Farahani, "Interaction between viscous varying modified cosmic Chaplygin gas and Tachyonic fluid", Astrophys. and Space Sci. 347 (2013) 209
\bibitem{P74}
J. Sadeghi, B. Pourhassan, M. Khurshudyan, H. Farahani, "Time-Dependent Density of Modified Cosmic Chaplygin
Gas with Variable Cosmological Constant in Non-Flat Universe", Int. J. Theor. Phys. 53 (2014) 911
\bibitem{P75 1402.2592}
E.O. Kahya, M. Khurshudyan, B. Pourhassan, R. Myrzakulov, A. Pasqua, "Higher order corrections of the extended Chaplygin gas cosmology with varying $G$ and $\Lambda$", Eur. Phys. J. C 75 (2015) 43 [arXiv:1402.2592 [gr-qc]]
\bibitem{P76 1405.0667}
B. Pourhassan, E.O. Kahya, "FRW cosmology with extended Chaplygin gas", ", Advances in High Energy Physics 2014 (2014) 231452 [arXiv:1405.0667 [gr-qc]]
\bibitem{a}
J. Sadeghi, H. Farahani, B. Pourhassan, "Interacting Holographic Extended Chaplygin Gas and Phantom Cosmology in the Light of BICEP2",  EPJP [arXiv:1412.1291 [gr-qc]]
\bibitem{b}
E.O Kahya, B. Pourhassan, "The universe dominated by the extended Chaplygin gas", MPLA [arXiv:1502.01189 [gr-qc]]
\bibitem{c}
E.O. Kahya, B. Pourhassan, "Observational constraints on the extended Chaplygin gas inflation", Astro Space Science 353 (2014) 677
\bibitem{P77 1212.2641}
M. Bouhmadi-Lopez, P. Chen, Y-C. Huang, Y-H. Lin, "Slow-Roll Inflation Preceded by a Topological Defect Phase � la Chaplygin Gas", Phys.Rev. D87 (2013) 10, 103513
\bibitem{P78 1303.5658}
R. Herrera, M. Olivares, N. Videla, "Intermediate-Generalized Chaplygin Gas inflationary universe model", Eur. Phys. J. C 73 (2013) 2295
\bibitem{P79 1310.4988}
S. del Campo, "Single-field inflation $\grave{a}$ la generalized Chaplygin gas", JCAP 11 (2013) 004 [arXiv:1310.4988 [astro-ph.CO]]
\bibitem{P80 1303.5075}
P.A.R. Ade et al. [Planck Collaboration], "Planck 2013 results. XV. CMB power spectra and likelihood", [arXiv:1303.5075 [astro-ph.CO]]
\bibitem{P81 1403.3985}
P.A.R. Ade et al. [BICEP2 Collaboration], "BICEP2 I: Detection Of B-mode Polarization at Degree Angular Scales", [arXiv:1403.3985 [astro.ph.CO]]
\bibitem{P82 1404.3683}
B.R. Dinda, S. Kumar, A.A. Sen, "Inflationary GCG + Phantom DE in the light of Planck and BICEP2", [arXiv:1404.3683 [astro-ph.CO]]
\bibitem{P83 1211.3518}
H.B. Benaoum, "Modified Chaplygin Gas Cosmology", Advances in High Energy Physics 2012 (2012) 357802
\bibitem{P84 1312.0779}
J. Lu, L. Xu, Y. Wu, M. Liu, "Reduced modified Chaplygin gas cosmology", [arXiv:1312.0779 [astro-ph.CO]]
\bibitem{P85 1004.3364}
J. Lu, L. Xu, J. Li, B. Chang, Y. Gui, H. Liu, "Constraints on modified Chaplygin gas from recent observations and a comparison of its status with other models", Phys. Lett. B 662 (2008) 87
\bibitem{P87 1105.1870}
J. Lu, L. Xu, Y. Wu, M. Liu, "Combined constraints on modified Chaplygin gas model from cosmological observed data: Markov Chain Monte Carlo approach", Gen. Rel. Grav. 43 (2011) 819
\bibitem{P91 1306.4808}
B.C. Paul, P. Thakur, "Observational Constraints on Modified Chaplygin Gas from Cosmic Growth", JCAP 1311 (2013) 052
\bibitem{P93 1205.3768}
D. Panigrahi, S. Chatterjee, "FRW type of cosmology with a Chaplygin gas", Int. J. Mod. Phys. D 21 (2012) 1250079
\bibitem{P94 1310.7167}
T. Harko, F.S.N. Lobo, M.K. Mak, "Arbitrary scalar field and quintessence cosmological models", Eur. Phys. J. C 74 (2014) 2784
\bibitem{1502.01589}
P.A.R. Ade et al. [Planck Collaboration], "Planck 2015 results. XIII. Cosmological Parameters", [arXiv:1502.01589]
\bibitem{P99 0905.2281}
P. Thakur, S. Ghose, B.C.Paul, "Modified Chaplygin Gas and Constraints on its B parameter from CDM and UDME Cosmological models", Mon. Not. R. Astron. Soc. 397 (2009) 1935
\bibitem{P57 1401.8002}
H. B. Benaoum, Modified Chaplygin Gas Cosmology with Bulk Viscosity in D Dimensions, Int. J. Mod. Phys. D 23, 1450082 (2014) [arXiv:1401.8002].
\bibitem{1405.7681}
J. Lee, T-H. Lee, P. Oh, J. Overduin, "Cosmological Coincidence without Fine Tuning", Phys. Rev. D 90 (2014) 123003 [arXiv:1405.7681 [hep-th]]
\end{thebibliography}
\end{document}